# Detecting quantum steering in networks

Ming-Xiao Li,[1,2,*] Yuqi Li,[3,*] Ya Xi,[4,†] Chang-Yue Zhang,[5] Ying-Zheng Wang,[1,2]
Rui-Bin Xu,[1,2] Shao-Ming Fei,[6] and Zhu-Jun Zheng[1,2,‡]

[1]*School of Mathematics, South China University of Technology, Guangzhou 510640, People's Republic of China*
[2]*Laboratory of Quantum Science and Engineering, South China University of Technology, Guangzhou 510641, People's Republic of China*
[3]*College of Mathematics Science, Harbin Engineering University, Harbin 150001, Heilongjiang, People's Republic of China*
[4]*Department of Mathematics and Date Sciences, Zhongkai University of Agriculture and Engineering,
Guangzhou 510225, People's Republic of China*
[5]*Department of Mathematics, College of Information Science and Technology, Jinan University,
Guangzhou 510632, People's Republic of China*
[6]*School of Mathematical Sciences, Capital Normal University, Beijing 100048, China*

Quantum networks promise an unprecedented leap in semi-device-independent communication and security by capitalizing on quantum steering. However, current methods for assessing quantum network steering are constrained to specific cases. In this work, we introduce the network-Clauser-Horn-Shimony-Holt-like inequality for investigating network steering independent of entanglement source characteristics. We employ this inequality to detect full network steering in both single-node and multinode repeater networks and assess the tolerance of various noise models. Under a specific noise model, our method is used to compute the bound of semi-device-independent communication distance. Through case studies, we also demonstrate that our method, as a semi-device-independent entanglement witness, is more suitable for settings requiring Bell measurements compared to network Bell inequality. These findings open new avenues for broader ways to detect quantum steering independent of entanglement sources.

DOI: 10.1103/PhysRevA.111.012624

## I. INTRODUCTION

Quantum networks play crucial roles in the development of quantum technologies [1–3], particularly in quantum communication and machine learning. Quantum correlations [4–11] in networks may offer better security and efficiency in information processing than the cases with classical correlations. Due to the practical challenges of decoherence, detecting quantum correlations among communicators is significant to verify the usability of quantum networks [12].

Widely employed techniques for detecting quantum nonlocalities, including quantum tomography and entanglement witnesses, presuppose the reliability of the measurement instruments, giving rise to considerable constraints. The device-independent (DI) methodology circumvents such limitations by leveraging statistical measurement outcomes to verify the violation of Bell inequalities [13,14]. The device-independent verification confers superior security and drives advancements in quantum communication, particularly in quantum networks and quantum key distribution (QKD) [15–17]. However, the implementation of device-independent communication poses significant technical challenges. For example, theoretical frameworks for DI-QKD demand a minimum system detection efficiency threshold of 90%, a benchmark that currently exceeds the capabilities of current technology [17]. Since the semi-device-independent (SDI) communication [18–20] is recognized for its potential to bridge the gap between information security and practicality, research on the detection of quantum steering has been of great interest.

Quantum steering, an asymmetric quantum correlation, enables one party to "steer" the other through local measurements [21–36]. Beyond standard steering, significant advances have been made in detecting multipartite unknown states' steering and repeater networks' steering [37–45]. In these cases, some parties employ trusted measurements, while other counterparts conduct "black box" measurements that are untrusted or uncharacterized. For instance, Ref. [39] introduces an semidefinite programming (SDP) method for detecting steering in multipartite systems, while Ref. [41] presents specific cases where steering detection in quantum networks depends on the source device's properties. These studies pave the way for the creation of hybrid trust-network communication systems better suited for real-world scenarios. These results motivate us to ask if a more universal method for detecting quantum steering can be established even in a fundamental repeater network with single relay node, and if such a method is applicable to longer-distance communication involving multiple nodes in the network.

In this work, we address the challenge by formulating a network-Clauser-Horn-Shimony-Holt (CHSH)-like inequality based on statistical measurement outcomes. This inequality function could be a semi-device-independent entanglement witness and serve as a critical tool for verifying the violations of the network local hidden state (NLHS) model, thereby assessing the steerability of quantum networks. We further demonstrate its robustness to various noise models in relay networks, confirming its potential in practical long-distance communication. Through the random generation of quantum states, we identify quantum networks where our method successfully detects network steering, even in cases where the entanglement criterion does not apply. When it comes to multinode scenarios, we generalize the inequality and conduct a comparative analysis to explore how noise tolerance varies with the increasing network nodes.

---

*These authors contributed equally to this work.
†Contact author: 960092758@qq.com
‡Contact author: zhengzj@scut.edu.cn

## II. PRELIMINARIES

In this section, we lay the groundwork for results presented in this work by introducing the notations and examining the framework of quantum network steering.

Quantum steering is a kind of quantum correlation that allows one observer to manipulate the conditional state of a spatially separated one through local measurements on their own system, defying explanation by any local hidden state (LHS) model. Consider a scenario where Alice generates a random variable $x \in \{1, \ldots, m\}$ to select a measurement from a positive operator-valued measure (POVM) set $\{M^A_{a|x}\}_{a=1,\ldots,n}$. She applies the measurement on the subsystem $A$ of a bipartite state $\rho^{AB}$ associated with the systems $A$ and $B$. Each POVM gives rise to $n$ possible measurement outcomes $a$. Upon performing the measurement $M^A_{a|x}$ on her subsystem, the resulting unnormalized conditional state of Bob's subsystem is given by $\widetilde{\rho}^B_{a|x} = \text{Tr}_A[(M^A_{a|x} \otimes \mathbb{1}^B)\rho^{AB}]$, which leads to a collection of the conditional states $\{\widetilde{\rho}^B_{a|x}\}$, termed as the assemblage. The correlations between Alice and Bob can be explained by an LHS model if each element in this assemblage admits a decomposition of the form [21,22],

$$\widetilde{\rho}^B_{a|x} = \sum_{\lambda} p(\lambda) p(a|x, \lambda) \sigma^B_{\lambda}, \quad (1)$$

where $\lambda$ is the hidden variable with the probability distribution $p(\lambda)$, $p(a|x, \lambda)$ is the local response function of Alice, and $\sigma_\lambda$ are given hidden states. A state $\rho^{AB}$ admitting an LHS model is said to be unsteerable from Alice to Bob. Conversely, if there exists a set of measurements that prevents $\widetilde{\rho}^B_{a|x}$ from the decomposition described by Eq. (1), we classify $\rho^{AB}$ as steerable from Alice to Bob. It is worth stressing that the reduced state of Bob's system, $\rho^B = \sum_a \widetilde{\rho}^B_{a|x}$, is inherently independent of Alice's measurement choices, allowing the steering to be detected even when Alice is untrusted. This property underscores its potential as a semi-device-independent approach for entanglement witness and a valuable resource for constructing semi-device-independent networks.

Long-distance and large-scale quantum communication necessitates relay nodes for entanglement swapping. Local measurements $M^{CC'}_c$ on the intermediate node Charlie establish correlations between Alice and Bob:

$$\sigma^{AB}_c = \text{Tr}_{CC'}\left[(\mathbb{1}^A \otimes M^{CC'}_c \otimes \mathbb{1}^C)\rho^{AC} \otimes \rho^{BC'}\right]. \quad (2)$$

Similar to quantum state steering, the postmeasurement state $\sigma^{AB}_c$ is characterized by an NLHS model, signifying the network is unsteerable if it can be decomposed in the following form,

$$\sigma^{AB}_c = \sum_{\beta, \gamma} p(\beta) p(\gamma) p(c \mid \beta, \gamma) \sigma^A_\beta \otimes \sigma^B_\gamma. \quad (3)$$

However, unlike the quantum state steering which necessitates multiple measurements, the network framework permits even a single fixed measurement $M^{CC'}_c$ to expose a violation of the NLHS model and confirm the network steerability. This framework offers a distinct practical advantage, closely aligning with real-world scenarios.

Building on the principles of CHSH, a renowned method for detecting quantum nonlocality, quantum steering can similarly be detected by optimizing measurement strategies and comparing experimental outcomes with the predictive limitations of the LHS model. These statistical results are captured by the CHSH-like inequalities [27],

$$\sqrt{\langle (A_1 + A_2)B_1 \rangle^2 + \langle (A_1 + A_2)B_2 \rangle^2}$$
$$+ \sqrt{\langle (A_1 - A_2)B_1 \rangle^2 + \langle (A_1 - A_2)B_2 \rangle^2} > 2, \quad (4)$$

where $A_i$ and $B_i$ are measurements performed by Alice and Bob, respectively. The only constraint on the measurement strategy is that $B_i$ are mutually unbiased measurements.

## III. DETECTING NETWORK QUANTUM STEERING WITH THREE PARTIES

The detection of network steerability is generally limited to specific conditions [41]. For a single-node network, a measurement set $\{M^C_{c|z}\}_{z=1,\ldots,d}$ may confirm the steerability of an entangled state $\rho^{AC}$. If another state $\rho^{BC'} = \sum_z \frac{1}{d}|z\rangle\langle z| \otimes |z\rangle\langle z|$ is taken into account, along with measurements $M^{CC'}_c = \sum_{z'} |z'\rangle\langle z'|^C \otimes M^{C'}_{c|z'}$, it verifies network steering. These scenarios impose strict requirements on communication parties. Here, we show a network-CHSH-like (NCHSH-like) inequality as a more general detection method independent of prior assumptions about the quantum sources.

### A. Network-CHSH-like inequality

In order to construct the NCHSH-like inequality, we introduce elegant joint measurements (EJMs) that can be implemented with a two-qubit quantum circuit [40,46]. EJMs are crucial for entanglement swapping and have diverse applications, ranging from superconducting quantum processors to photonic quantum walks. To define the generalized EJMs, we first introduce four pure states $|\vec{m}_c\rangle$ on the Bloch sphere, each pointing towards one of the four vertices,

$$\vec{m}_1 = (+1, +1, +1), \quad \vec{m}_2 = (+1, -1, -1),$$
$$\vec{m}_3 = (-1, +1, -1), \quad \vec{m}_4 = (-1, -1, +1), \quad (5)$$

as well as the states $|-\vec{m}_c\rangle$ with the antipodal direction. Using cylindrical coordinates, we write these tetrahedron vertices as

$$\vec{m}_c = \sqrt{3}\left(\sqrt{1-\eta_c^2}\cos\phi_c, \sqrt{1-\eta_c^2}\sin\phi_c, \eta_c\right).$$

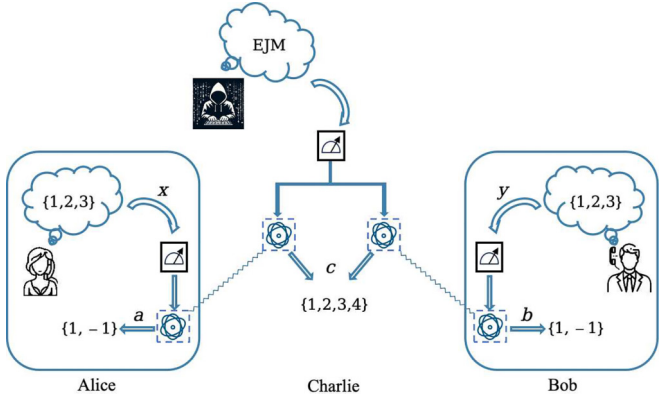

FIG. 1. Three-party network quantum steering scenario. Charlie shares a pair of entangled state $\rho_{AC}$ with Alice and another entangled state $\rho_{BC'}$ with Bob. Charlie performs a fixed uncharacterized measurement with output $c \in \{1, 2, 3, 4\}$, while Alice and Bob perform characterized measurements $x, y \in \{1, 2, 3\}$ with output $a, b \in \{+1, -1\}$, respectively.

Then the corresponding states are of the form

$$|\pm \vec{m}_c\rangle = \sqrt{\frac{1 \pm \eta_c}{2}} e^{i\phi_c/2}|0\rangle \pm \sqrt{\frac{1 \mp \eta_c}{2}} e^{-i\phi_c/2}|1\rangle. \quad (6)$$

The generalized EJMs are then defined by

$$|\Phi_c^\theta\rangle = \frac{\sqrt{3} + e^{i\theta}}{2\sqrt{2}}|\vec{m}_c, -\vec{m}_c\rangle + \frac{\sqrt{3} - e^{i\theta}}{2\sqrt{2}}|-\vec{m}_c, \vec{m}_c\rangle. \quad (7)$$

We obtain the EJM introduced in Ref. [47] when $\theta = 0$ and the Bell state measurement up to local unitaries $U_1 \otimes U_2 = \mathbb{1} \otimes e^{(2\pi i/3)[(\sigma_1+\sigma_2+\sigma_3)/\sqrt{3}]}$ when $\theta = \pi/2$.

In our framework, as shown in Fig. 1, we introduce an SDI repeater network with a hierarchical trust structure: Alice and Bob, representing the laboratory-based devices and end users, respectively, are trusted, while other parties like the relay node, Charlie, are untrusted due to potential adversarial behavior. Charlie performs the EJM defined in Eq. (7) on his qubits, with outcomes $c = (c^1, c^2, c^3) \in \{\vec{m}_i\}_{i=1,\ldots,4}$ corresponding to the vectors in Eq. (5). This induces quantum correlations between qubits $A$ and $B$. Alice and Bob independently measure their spin observables along three orthogonal $\{\vec{s_{x(y)}}\}_{x(y)=1,2,3}$ axes, with inputs $x, y \in \{1, 2, 3\}$ and outcomes $a, b \in \{\pm 1\}$. The correlations among the three measurements are characterized by the expectation value:

$$\langle A_x C^k B_y \rangle = \sum_{a,c,b} ac^k b p(a, c, b|x, y), \quad (8)$$

where $p(a, c, b|x, y)$ is the conditional probability distribution, and $c^k$ represents the $k$th element in the array $c = (c^1, c^2, c^3)$. Under the NLHS model in Eq. (3), $p(a, c, b|x, y)$ can be expressed by

$$p(a, c, b|x, y) = \sum_{\beta, \gamma} p(\beta) p(\gamma) p(c|\beta, \gamma) p(a|x, \sigma_\beta^A) p(b|y, \sigma_\gamma^B). \quad (9)$$

To set up the main theorem, let us first introduce a useful result of a one-body expectation.

*Lemma 1.* For any pure state, consider observables $\{A_x\}_{x=1,2,3}$ measured via spin along three orthogonal $\vec{s_x}$ axes. The expectation values satisfy

$$\langle A_1 \rangle^2 + \langle A_2 \rangle^2 + \langle A_3 \rangle^2 = 1. \quad (10)$$

The proof of the theorem is provided in Appendix A. Note that the measurements used by Alice and Bob are mutually unbiased. We present the method for detecting steerability in repeater networks within the scenario considered below.

*Theorem 1.* If the full repeater network admits an NLHS model, then the correlations among the three parties are bounded by the following NCHSH-like inequality,

$$\sqrt{\sum_{x,y=1}^{3} \langle A_x(C^2 + C^3)B_y \rangle^2} + \sqrt{\sum_{x,y=1}^{3} \langle A_x(C^1 - C^2)B_y \rangle^2}$$
$$+ \sqrt{\sum_{x,y=1}^{3} \langle A_x(C^1 - C^3)B_y \rangle^2} \leqslant 2. \quad (11)$$

The proof of the theorem is provided in Appendix B. This inequality constitutes a necessary condition for a quantum network to admit an NLHS model. Any violation of the inequality implies the steerability of the network. Accordingly, the NCHSH-like inequalities serve as semi-device-independent entanglement witnesses.

### B. Application in different noise models

We illustrate how our approach captures the effects of different noise types on networks. This allows for the assessment of whether a network can support information transmission under noisy environments. We simplify the measurement setup by considering that Alice and Bob independently measure the Pauli observables $\sigma_X$, $\sigma_Y$, and $\sigma_Z$ after Charlie's generalized EJMs on his systems.

We first analyze two entangled singlet states subjected to depolarizing noise:

$$\rho_D^i = (1 - v_i)|\phi\rangle\langle\phi| + v_i \frac{I}{4}, \quad i = 1, 2, \quad (12)$$

where $|\phi\rangle = (|01\rangle - |10\rangle)/\sqrt{2}$, and $I$ is the identity matrix of $4 \times 4$. The resulting three-body correlations are affected by the measurement settings $\theta$ and the noise parameters $v_i$ [40]:

$$\langle A_x C^k B_y \rangle$$
$$= \begin{cases} -\frac{(1-v_1)(1-v_2)}{2}(1 + \sin\theta) & \text{if } xzy \in \{123, 231, 312\}, \\ -\frac{(1-v_1)(1-v_2)}{2}(1 - \sin\theta) & \text{if } xzy \in \{132, 213, 321\}, \\ 0 & \text{otherwise.} \end{cases} \quad (13)$$

Substituting it into Eq. (11) we get the left side of the inequality, $3(1 - v_1)(1 - v_2)\sqrt{1 + \sin^2\theta}$. When there is no noise ($v_i = 0$), the NCHSH-like inequality is always violated regardless of the measurement settings $\theta$. In contrast, the violation of the inequality under nonzero noise depends on both $v_i$ and $\theta$. For instance, at $\theta = \frac{\pi}{2}$, the inequality is violated if $(1 - v_1)(1 - v_2) > \sqrt{2}/3$, as shown in Fig. 2(a).

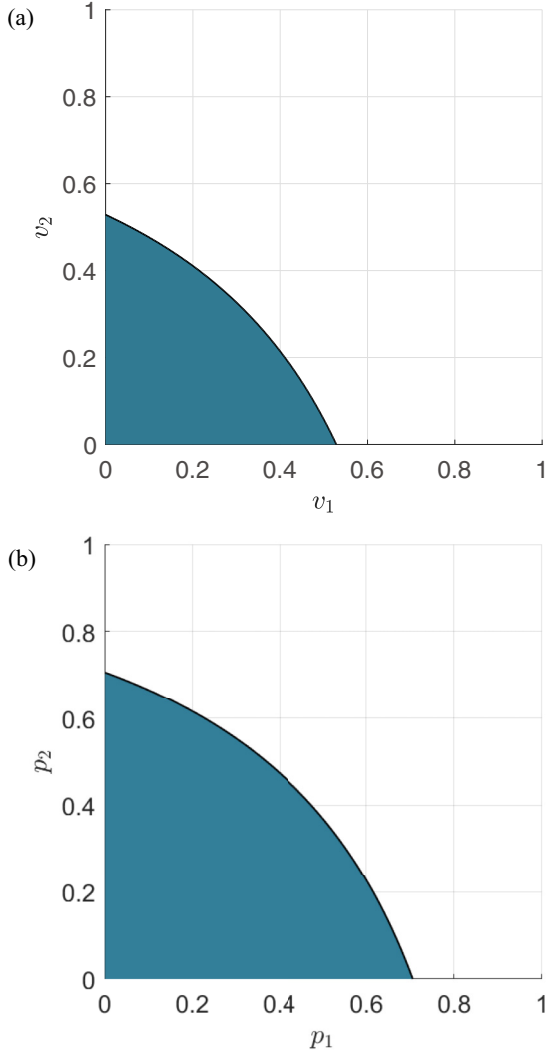

FIG. 2. Steerability of a three-party quantum network under different noises. Parameters $v_i$ and $p_i$ denote different noises of two sources. The green areas in the figures indicate the range of noise parameters in which the network exhibits steerability. Panel (a) is for $\theta = \frac{\pi}{2}$. $v_1$ and $v_2$ take values from 0 to 0.53. Panel (b) is for the amplititude damping noise. $p_1$ and $p_2$ take values from 0 to 0.71.

For the second scenario we consider the amplitude damping noise within the network, in which the shared states of the two pairs are given by

$$\rho_{AD} = \sum_j (I \otimes K_j)|\phi\rangle\langle\phi|(I \otimes K_j)^\dagger, \quad (14)$$

with Kraus operators $K_j$ defined as

$$K_1 = \begin{bmatrix} 0 & \sqrt{p_i} \\ 0 & 0 \end{bmatrix}, \quad K_2 = \begin{bmatrix} 1 & 0 \\ 0 & \sqrt{1-p_i} \end{bmatrix}. \quad (15)$$

Setting $\theta = \frac{\pi}{2}$ we get the correlations $\langle A_1 C^2 B_3 \rangle = (p_2 - 1)\sqrt{1-p_1}$, $\langle A_3 C^1 B_2 \rangle = (p_1 - 1)\sqrt{1-p_2}$, and $\langle A_2 C^3 B_1 \rangle = -\sqrt{(1-p_1)(1-p_2)}$, with all the other ones vanishing. Substituting these into Eq. (11), we obtain that the inequality is violated when

$$\sqrt{(1-p_1)(1-p_2)(2-p_1-p_2)}$$
$$+ \sqrt{(1-p_1)(1-p_2)(2-p_2)}$$
$$+ \sqrt{(1-p_1)(1-p_2)(2-p_1)} > 2. \quad (16)$$

The permissible range of the noise parameters for inequality to be violated under amplitude damping noise is depicted in the green region of Fig. 2(b).

### C. Distance of repeater networks

The entanglement distribution is crucial for quantum communication, which enables the creation of entangled states between distant parties and underpins the quantum repeater networks. In this process, the relay node Charlie prepares two entangled pairs and sends one half to Alice and Bob, respectively. Establishing reliable entanglement over long distances needs to overcome decoherence and other noise effects. Optimal repeater placement is thus vital for constructing effective quantum networks, so as to maintain reliable quantum correlations with minimized resource waste by optimizing the effective distance [48,49]. Our method can be used to compute the feasible entanglement distribution distances under various noises, thereby determining the maximum transmission range of repeater networks with a single node.

We consider the distribution of entanglement by Charlie through a depolarization channel of length $l$ with the noise parameter $\alpha$ given by

$$\mathcal{N}_l(\cdot) = e^{-\alpha l}(\cdot) + (1 - e^{-\alpha l})\frac{\mathbb{I}}{2}. \quad (17)$$

Consider a scenario in which Charlie transmits one party of a singlet state through depolarizing channels to Alice and Bob, with the lengths of the respective channels denoted as $l_1$ and $l_2$. Charlie performs a generalized EJM with $\theta = \pi/2$, while Alice and Bob carry out three Pauli measurements. The resulting correlations are $\langle A_1 C^2 B_3 \rangle = \langle A_3 C^1 B_2 \rangle = \langle A_2 C^3 B_1 \rangle = -e^{-\alpha(l_1+l_2)}$, with all other expectations vanishing. The inequality (11) is violated if $l_1 + l_2 < \frac{\ln(3/\sqrt{2})}{\alpha}$, indicating the presence of quantum steering (see Fig. 3).

### D. Comparison with bilocal Bell inequality

In Ref. [40] the bilocal inequality was employed to characterize the violation of the quantum network nonlocality. Nonlocality serves as a device-independent entanglement and is particularly important in adversarial scenarios such as device-independent quantum key distribution. However, the violation of a Bell inequality usually demands exceptionally high correlations between the parties, which necessitate minimal noise, near-perfect detectors, and high-quality entangled states. We have developed a method that reduces the need for device independence in quantum networks. Our inequality in Theorem 1 serves as a semi-device-independent resource. Here, we present numerical experiments and comparisons between NCHSH-like inequality and the bilocal inequality. Let

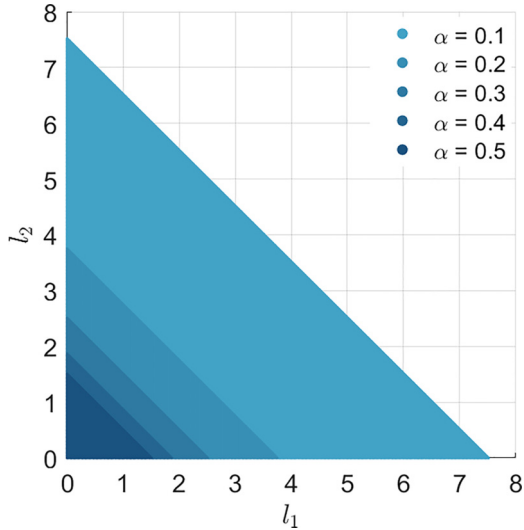

FIG. 3. Channel length ranges of three-party steerable networks under white noise. The lengths of depolarizing channels are denoted by $l_1$ and $l_2$. The blue areas show the ranges of channel lengths for which the quantum network exhibits steerability through the depolarizing channels for damping parameters $\alpha = 0.1, 0.2, 0.3, 0.4$, and $0.5$.

us begin with the construction of the following quantities:

$$S = \sum_{y=z}\langle B_y C^k\rangle - \sum_{x=z}\langle A_x C^k\rangle,$$

$$T = \sum_{x\neq y\neq z\neq x}\langle A_x C^k B_y\rangle, \quad Z = \max(\mathcal{C}_{\text{other}}), \quad (18)$$

where $\mathcal{C}_{\text{other}} = \{|\langle A_x\rangle|, |\langle A_x C^k\rangle|, \ldots, |\langle A_x C^k B_y\rangle|\}$ represents the set of absolute values of all one-, two-, and three-party correlators, excluding those appearing in the expressions for $S$ and $T$. Then, the network is bilocal if the following condition holds,

$$\mathcal{B} \equiv \frac{S}{3} - T \leqslant 3 + 5Z. \quad (19)$$

To enable a clear comparison, we consider the scenario with the state in Eq. (12), where the correlations depend on both the measurement settings $\theta$ and the noise parameters $v$. The three-body correlators are consistent with Eq. (13), while the one-body and two-body correlators are given by

$$\langle A_x\rangle = \langle C^k\rangle = \langle B_y\rangle = \langle A_x B_y\rangle = 0,$$

$$\langle A_x C^k\rangle = -\frac{(1-v_1)}{2}\cos\theta\,\delta_{x,z},$$

$$\langle B_y C^k\rangle = \frac{(1-v_2)}{2}\cos\theta\,\delta_{y,z}.$$

When the noise parameters are equal, i.e., $v_1 = v_2 = v$, the bilocality inequality is violated if $3(1-v)^2 + (1-v)\cos\theta > 3$. In contrast, the steering inequality is violated under the condition $3(1-v)^2\sqrt{1+\sin^2\theta} > 2$.

In Fig. 4, we show the bounds on the noise parameters required to violate the bilocality and steering inequalities under various measurement settings. Between these bounds, the network states are steerable, but may not exhibit nonlocality, and thus may not confer the expected quantum advantage in device-independent protocols. However, steering from the violation of the steering inequalities tolerates higher noise, especially when the Bell measurements at repeater nodes require a noise-free condition to violate the bilocality inequality. Therefore, for tasks that involve Bell measurements at repeater nodes, leveraging network steering as a resource for semi-device-independent communication is a more practical strategy, balancing feasibility with security considerations.

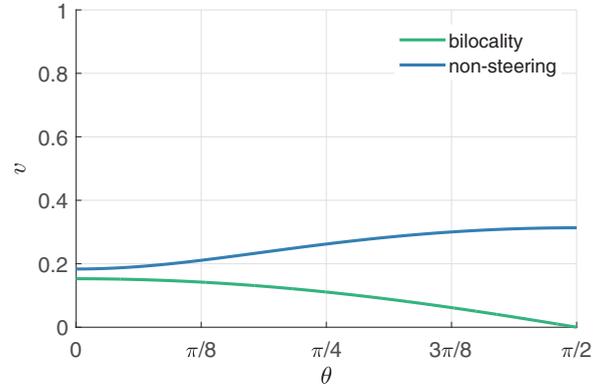

FIG. 4. Comparison of non-bilocality and steering for the case of two sources under the same white noise. The figure shows the ranges of white noise parameters in which the network exhibits steering and the Bell non-bilocality under different measurement parameters $\theta$. Below the red line the network demonstrates steering by the entanglement criterion. Below the blue line the network demonstrates steering by the NCHSH-like inequality.

### E. Comparison with entanglement criteria

To date, we have proposed the NCHSH-like inequality, which is particularly well-suited for scenarios where fully characterizing the density matrix of the quantum state is difficult. For instance, the combination of diverse noise sources, technical limitations, and instrumental imperfections in real-world communication systems significantly distorts quantum states, making precise characterization challenging. If complete and accurate tomography was achievable, enabling a full reconstruction of the density matrix, additional degrees of freedom would become accessible for evaluating network steering. In such cases, a direct approach would be to apply entanglement criteria to assess whether the postmeasurement state shared between Alice and Bob exhibits entanglement. If entanglement is observed, a violation of the NLSH model, as defined in Eq (3), would confirm network steerability.

This leads us to the following question: Does our method retain its significance when partial or complete characterization of the postmeasurement density matrix is feasible? To investigate this, we performed a rigorous comparison using the positive partial transpose (PPT) criterion in low-dimensional systems. Under this criterion, the presence of a negative eigenvalue in the partially transposed density matrix signifies entanglement. In qubit-qubit and qubit-qutrit systems, this criterion is both necessary and sufficient for entanglement detection.

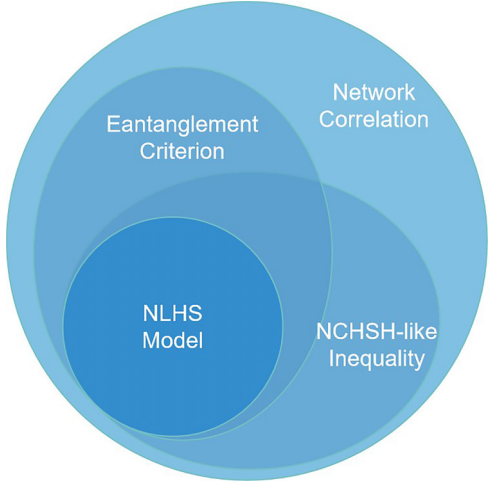

FIG. 5. The correlation between the entanglement criterion and the NCHSH-like inequality. Both of them serve as necessary conditions for the NLHS model. The region outside the ellipse labeled "Inequality" indicates where the inequality successfully certifies network steering, while the region outside the ellipse labeled "Entanglement Criterion" represents where the entanglement criterion certifies network steering.

The answer is affirmative: we have verified the intersection relationship between the NCHSH inequality and the PPT criterion, as shown in Fig. 5. We randomly generated 100 quantum states for $\rho^{AC}$ and $\rho^{BC'}$, respectively, with Charlie performing a generalized EJM set to $\theta = \pi/2$. Among the cases where the PPT criterion failed to certify network steering, the NCHSH-like inequality succeeded in detecting steering for 27 instances. Three representative examples are detailed in Appendix C, with a thorough analysis of the first example provided. This implies that combining both methods will provide a more effective means of detecting network steering.

## IV. DETECTION OF DUAL-NODE CHAINED NETWORK STEERING

Adding nodes is essential for large-scale networks, as it enhances coverage and extends signal reach by enabling internode communication. In this section, we introduce an additional node, David, with the same privileges and configuration as the relay node Charlie (see Fig. 6). This improvement builds on basic relay configurations and introduces a dual-node setup that facilitates connectivity. We analyze the steerability of a repeater network with a dual-node chain topology by generalizing the Network-CHSH-like inequality.

### A. Dual-node network-CHSH-like inequality

We commence by delineating the NLHS model for a dual-node network. Specifically, denote $\sigma_{c,d}^{AB}$ the postmeasurement state resulting from the execution of measurements $M_c^{CC'}$ and $M_d^{DD'}$. The network adheres to an NLHS model if the subnormalized states $\sigma_{c,d}^{AB}$ can be expressed in the following form,

$$\sigma_{c,d}^{AB} = \sum_{\alpha,\beta,\gamma} p(\alpha)p(\beta)p(\gamma)p(c \mid \alpha, \beta)p(d \mid \beta, \gamma)\sigma_\alpha^A \otimes \sigma_\gamma^B. \quad (20)$$

The four-body correlators are given by [see the detailed derivations Eq. (B2) given in Appendix B]

$$\langle A_x C^k D^l B_y \rangle = \sum_{a,c,d,b} a c^k d^l b\, p(a, c, d, b | x, y)$$
$$= \sum_{c,d,\xi,\zeta} p(c, d, \xi, \zeta) c^k d^l \langle A_x \rangle_{|\phi_\xi\rangle\langle\phi_\xi|} \langle B_y \rangle_{|\psi_\zeta\rangle\langle\psi_\zeta|}. \quad (21)$$

Leveraging the methodology from the proof of Theorem 1 in Appendix B, we extend below the NCHSH-like inequality to provide a broader framework for network steering analysis.

*Theorem 2.* If the full dual-node repeater network admits an NLHS model, then the correlations among the four

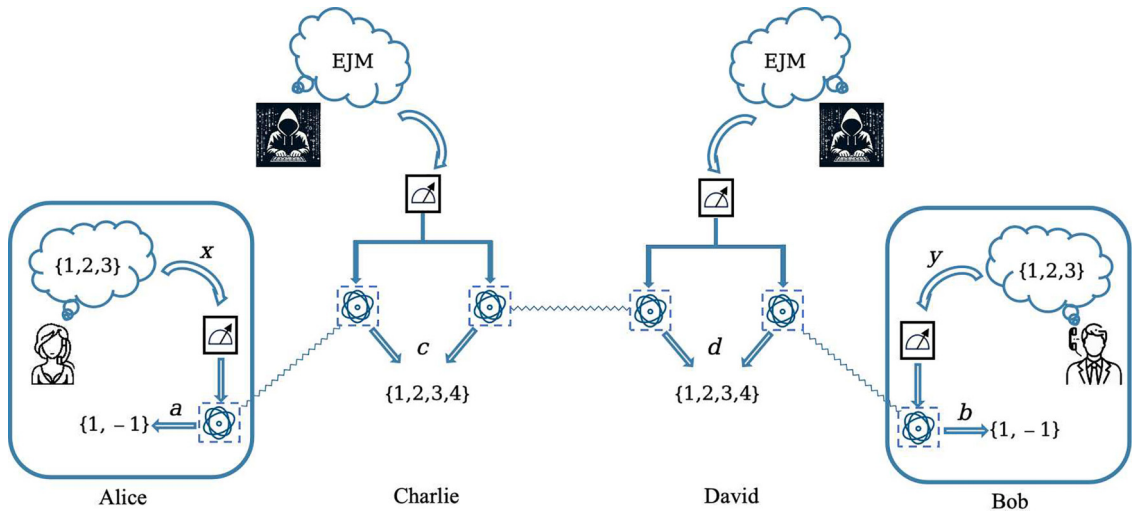

FIG. 6. Four-party chained network steering scenario. Alice and Bob perform characterized measurements $x, y \in \{1, 2, 3\}$ with outputs $a, b \in \{+1, -1\}$, respectively, while Charlie and David perform fixed uncharacterized measurements with outputs $c, d \in \{1, 2, 3, 4\}$, respectively.

parties are bounded by the following multipartite NCHSH-like inequality,

$$\sum_{\substack{1 \leqslant m < n \leqslant 3 \\ 1 \leqslant p < q \leqslant 3}} \sqrt{\sum_{x,y=1}^{3} \langle A_x [C^m + (-1)^m C^n][D^p + (-1)^p D^q] B_y \rangle^2} \leqslant 4. \quad (22)$$

The proof of the theorem is similar to that of Theorem 1. We provide a brief process in Appendix D. The equality establishes a necessary condition for the dual-node chained networks to admit an NLHS model. Any violation of the above inequality implies that the network is steerable.

### B. Application in different noise models

In the previous sections, we have analyzed the noise tolerance of the NCHSH-like inequalities in single-node repeater networks. Here, we extend the analysis to networks with additional nodes by using the same numerical setup. In this scenario, David, an untrusted intermediate node, performs the same measurement as Charlie with the setting $\theta = \frac{\pi}{2}$.

Under the depolarizing noise, the correlations are

$$\langle A_1 C^2 D^1 B_2 \rangle = (1 - v_1)(1 - v_2)(1 - v_3),$$
$$\langle A_2 C^3 D^2 B_3 \rangle = -(1 - v_1)(1 - v_2)(1 - v_3),$$
$$\langle A_3 C^1 D^3 B_1 \rangle = (1 - v_1)(1 - v_2)(1 - v_3), \quad (23)$$

with all other correlations vanishing. The maximum violation of the inequality (22) occurs in the absence of noise ($v = 0$), yielding a value of $6 + 3\sqrt{2}$. When the noise is present, the quantum steerability is maintained if $(1 - v_1)(1 - v_2)(1 - v_3) > \frac{4}{6 + 3\sqrt{2}}$, as shown in Fig. 7(a).

For amplitude damping noise, each source $S_i$ introduces damping characterized by the decay parameter $p_i$ ($i = 1, 2,$ and 3). The resulting correlations are

$$\langle A_1 C^2 D^1 B_2 \rangle = (1 - p_2)\sqrt{(1 - p_1)(1 - p_3)},$$
$$\langle A_2 C^3 D^2 B_3 \rangle = (p_3 - 1)\sqrt{(1 - p_1)(1 - p_2)},$$
$$\langle A_3 C^1 D^3 B_1 \rangle = (p_1 - 1)\sqrt{(1 - p_2)(1 - p_3)}. \quad (24)$$

Denote the three correlations listed above as $u_1$, $u_2$, and $u_3$, respectively. The violation condition of the inequality (22) becomes

$$\sqrt{u_1^2 + u_2^2} + \sqrt{u_1^2 + u_3^2} + \sqrt{u_2^2 + u_3^2}$$
$$+ 2(|u_1| + |u_2| + |u_3|) > 4. \quad (25)$$

This condition defines the steering region under amplitude damping noise, as depicted in Fig. 6(b), where the quadripartite network exhibits steering below the defined surface.

An intriguing observation from Figs. 2 and 7 is that, as the number of nodes increases, the tolerance of steering detection to noise not only persists but even improves across different noise types. This enhancement implies that multinode networks are more resilient to disturbances, highlighting the potential of steering as a more robust semi-device-independent resource in complex quantum networks.

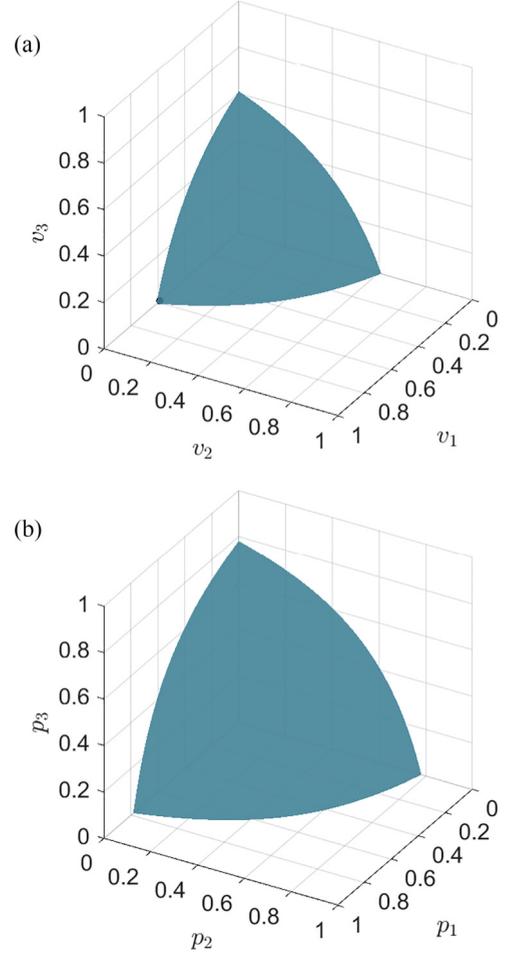

FIG. 7. Steerability of a four-party linear quantum network under different noise. Parameters $v_i$ and $p_i$ denote different noises of two sources. The area beneath the curved surface indicates the range of noise parameters within which the network exhibits steerability. Panel (a) is for the white noise with intersection coordinates (0.61,0,0), (0,0.61,0), and (0,0,0.61). Panel (b) is for the amplitude damping noise with intersection coordinates (0.80,0,0), (0,0.80,0), and (0,0,0.80).

## V. CONCLUSION

We have introduced the network-CHSH-like inequality, which gives a way to evaluate the quantum network steerability independent of entanglement source characteristics. We examined the noise thresholds for the inequality violation across different models and found that these thresholds vary with the number of network nodes, even under the same noise conditions. Contrary to intuition, increasing the number of repeater nodes does not require stricter noise constraints to detect the network steerability. We have also demonstrated the applications of the inequality in determining the effective distance for semi-device-independent communication under specific noise models. Compared to the entanglement criterion, which serves as another sufficient condition for network steering, our method successfully demonstrates network steering in certain quantum networks where the entanglement criterion fails.

As a semi-device-independent entanglement witness, our approach outperforms the device-independent methods in terms of detection range. From a communication security standpoint, while device-independent protocols offer the highest security levels, they require exceptionally low noise conditions. Specifically, we have shown that violating the bilocal inequality with Bell measurements at repeater nodes is only feasible without white noise. In contrast, our method tolerates more noise, offering a practical compromise between security and feasibility.

Our research provides a foundation for detecting network steering, yet large-scale quantum networks involve more complex topologies with numerous communication parties and relay nodes. Our studies may also shed new light on investigating other quantum network configurations or information-theoretic tasks, such as the star-shaped or tree-shaped networks.

## ACKNOWLEDGMENTS

We express our sincere gratitude to Mao-Sheng Li and Jun-Jing Xing for their invaluable support and insightful discussions that significantly contributed to the completion of this work. This work is supported by the Key Lab of Guangzhou for Quantum Precision Measurement under Grant No. 202201000010, the National Natural Science Foundation of China under Grants No. 12371458, No. 12075159, and No. 12171044, and the specific research fund of the Innovation Platform for Academicians of Hainan Province.

## APPENDIX A: PROOF OF LEMMA 1

*Proof.* Any pure state $\rho$ can be represented as a point on the Bloch sphere, $\rho = \frac{1}{2}(I + \vec{r} \cdot \vec{\sigma})$, with $|\vec{r}| = 1$. The observables $A_x$ can be expressed as linear combinations of Pauli matrices, $A_x = \sum_{j=1}^{3} s_{xj}\sigma_j$, with respect to three orthogonal vectors $(s_{x1}, s_{x2}, s_{x3})$. The summation of the squared expectation values of the observables is given by

$$\langle A_1 \rangle^2 + \langle A_2 \rangle^2 + \langle A_3 \rangle^2 = \text{Tr}(A_1\rho)^2 + \text{Tr}(A_2\rho)^2 + \text{Tr}(A_3\rho)^2$$

$$= \sum_{x=1}^{3} \left( \sum_{j=1}^{3} s_{xj} r_j \right)^2$$

$$= \sum_{j=1}^{3} \sum_{x=1}^{3} (s_{xj} r_j)^2 + 2 \left( \sum_{x=1}^{3} (r_1 r_2 s_{x1} s_{x2} + r_1 r_3 s_{x1} s_{x3} + r_2 r_3 s_{x2} s_{x3}) \right). \tag{A1}$$

Due to the orthonormality of the vectors, we have $\sum_{x=1}^{3} s_{xj_1} s_{xj_2} = \delta_{j_1 j_2}$. Therefore, $\langle A_1 \rangle^2 + \langle A_2 \rangle^2 + \langle A_3 \rangle^2 = 1$. ∎

## APPENDIX B: PROOF OF THEOREM 1

*Proof.* Let the states $\sigma_\beta$ and $\sigma_\gamma$ have the following pure state decompositions, $\sigma_\beta = \sum_\xi p(\xi|\beta) |\phi_\xi\rangle\langle\phi_\xi|$ and $\sigma_\gamma = \sum_\zeta p(\zeta|\gamma) |\psi_\zeta\rangle\langle\psi_\zeta|$. Notably, $\xi$ is independent of $\gamma$; $\zeta$ is independent of $\beta$; and $c$, $\xi$, and $\zeta$ are independent of each other. Hence, we have $p(c|\beta, \gamma) p(\xi|\beta) p(\zeta|\gamma) = p(c, \xi, \zeta|\beta, \gamma)$. Then, for a probability distribution which has an NLHS model, we have

$$p(a, c, b|x, y) = \sum_{\beta, \gamma} \sum_{\xi, \zeta} p(\beta) p(\gamma) p(c, \xi, \zeta|\beta, \gamma) p(a|x, |\phi_\xi\rangle\langle\phi_\xi|) p(b|y, |\psi_\zeta\rangle\langle\psi_\zeta|)$$

$$= \sum_{\xi, \zeta} p(c, \xi, \zeta) p(a|x, |\phi_\xi\rangle\langle\phi_\xi|) p(b|y, |\psi_\zeta\rangle\langle\psi_\zeta|). \tag{B1}$$

Therefore, the expectation defined in Eq. (8) can be expressed as

$$\langle A_x C^k B_y \rangle = \sum_{a,c,b} ac^k b \, p(a, c, b|x, y)$$

$$= \sum_{a,c,b} \left( ac^k b \sum_{\xi, \zeta} p(c, \xi, \zeta) p(a|x, |\phi_\xi\rangle\langle\phi_\xi|) p(b|y, |\psi_\zeta\rangle\langle\psi_\zeta|) \right)$$

$$= \sum_{c,\xi,\zeta} \left( p(c, \xi, \zeta) c^k \sum_a a \, p(a|x, |\phi_\xi\rangle\langle\phi_\xi|) \sum_b b \, p(b|y, |\psi_\zeta\rangle\langle\psi_\zeta|) \right)$$

$$= \sum_{c,\xi,\zeta} p(c, \xi, \zeta) c^k \langle A_x \rangle_{|\phi_\xi\rangle\langle\phi_\xi|} \langle B_y \rangle_{|\psi_\zeta\rangle\langle\psi_\zeta|}, \tag{B2}$$

where $\langle A_x \rangle_{|\phi_\xi\rangle\langle\phi_\xi|}$ and $\langle B_y \rangle_{|\psi_\zeta\rangle\langle\psi_\zeta|}$ are the expectations of measurements $A_x$ on the state $|\phi_\xi\rangle\langle\phi_\xi|$ and $B_y$ on the state $|\psi_\zeta\rangle\langle\psi_\zeta|$, respectively.

TABLE I. Expectations of the NLHS model in repeater networks.

| Expectations | $c = (1, 1, 1)$ | $c = (1, -1, -1)$ | $c = (-1, 1, -1)$ | $c = (-1, -1, 1)$ |
|---|---|---|---|---|
| $\langle A_x C^1 B_y \rangle$ | $\langle A_x \rangle \langle B_y \rangle$ | $\langle A_x \rangle \langle B_y \rangle$ | $-\langle A_x \rangle \langle B_y \rangle$ | $-\langle A_x \rangle \langle B_y \rangle$ |
| $\langle A_x C^2 B_y \rangle$ | $\langle A_x \rangle \langle B_y \rangle$ | $-\langle A_x \rangle \langle B_y \rangle$ | $\langle A_x \rangle \langle B_y \rangle$ | $-\langle A_x \rangle \langle B_y \rangle$ |
| $\langle A_x C^3 B_y \rangle$ | $\langle A_x \rangle \langle B_y \rangle$ | $-\langle A_x \rangle \langle B_y \rangle$ | $-\langle A_x \rangle \langle B_y \rangle$ | $\langle A_x \rangle \langle B_y \rangle$ |

Since from Lemma 1, when a set of mutually unbiased measurements is performed on a pure qubit state, the sum of the squares of the expectations is equal to 1, we can assign the following,

$$\langle A_1 \rangle_{|\phi_\xi\rangle\langle\phi_\xi|} = \cos\theta_1 \cos\theta_2, \quad \langle B_1 \rangle_{|\psi_\varsigma\rangle\langle\psi_\varsigma|} = \cos\varphi_1 \cos\varphi_2,$$
$$\langle A_2 \rangle_{|\phi_\xi\rangle\langle\phi_\xi|} = \cos\theta_1 \sin\theta_2, \quad \langle B_2 \rangle_{|\psi_\varsigma\rangle\langle\psi_\varsigma|} = \cos\varphi_1 \sin\varphi_2,$$
$$\langle A_3 \rangle_{|\phi_\xi\rangle\langle\phi_\xi|} = \sin\theta_1, \qquad \langle B_3 \rangle_{|\psi_\varsigma\rangle\langle\psi_\varsigma|} = \sin\varphi_1. \tag{B3}$$

Consequently, the 27 expectations are presented in Table I (where $\langle A_x \rangle_{|\phi_\xi\rangle\langle\phi_\xi|}$ and $\langle B_y \rangle_{|\psi_\varsigma\rangle\langle\psi_\varsigma|}$ are abbreviated as $\langle A_x \rangle$ and $\langle B_y \rangle$).

It is evident that the last column is a convex combination of the first three columns. Therefore, we may focus on the first three columns. Then if a three-party network admits an NLHS model under the specified measurements, the corresponding 27 expectations must be a convex combination of the first three columns for some values of $\theta_1, \theta_2, \varphi_1$, and $\varphi_2$. To formalize this, we define the sets $\mathfrak{C}_1, \mathfrak{C}_2$, and $\mathfrak{C}_3$ as the convex hulls of the three sets of points in $\mathbb{R}^{27}$, with each parametrized by $\theta_1, \theta_2, \varphi_1$, and $\varphi_2$. Hence, the NLHS model resides within the convex hull $\mathfrak{C}$ of the union of $\mathfrak{C}_1, \mathfrak{C}_2$, and $\mathfrak{C}_3$,

$$\mathfrak{C} = \text{convex}(\mathfrak{C}_1 \cup \mathfrak{C}_2 \cup \mathfrak{C}_3). \tag{B4}$$

In order to better describe the region $\mathfrak{C}$, we define a new orthogonal basis in $\mathbb{R}^{27}$,

$$\vec{e}_i = (\ldots, 0, \underset{i\text{th}}{1}, 0, \ldots 0, \underset{i+9\text{th}}{1}, 0, \ldots, 0, \underset{i+18\text{th}}{1}, 0, \ldots),$$
$$\vec{e}_{i+9} = (\ldots, 0, \underset{i\text{th}}{1}, 0, \ldots 0, \underset{i+9\text{th}}{-1}, 0, \ldots, 0, \underset{i+18\text{th}}{-1}, 0, \ldots),$$
$$\vec{e}_{i+18} = (\ldots, 0, \underset{i\text{th}}{-1}, 0, \ldots 0, \underset{i+9\text{th}}{1}, 0, \ldots, 0, \underset{i+18\text{th}}{-1}, 0, \ldots),$$

where $1 \leqslant i \leqslant 9$. In terms of this basis, the boundaries of $\mathfrak{C}_1, \mathfrak{C}_2$, and $\mathfrak{C}_3$ are of the forms,

$$\partial\mathfrak{C}_1 = \cos\theta_1 \cos\theta_2 \cos\varphi_1 \cos\varphi_2 \vec{e}_1 + \cos\theta_1 \cos\theta_2 \cos\varphi_1 \sin\varphi_2 \vec{e}_2 + \cos\theta_1 \cos\theta_2 \sin\varphi_1 \vec{e}_3$$
$$+ \cos\theta_1 \sin\theta_2 \cos\varphi_1 \cos\varphi_2 \vec{e}_4 + \cos\theta_1 \sin\theta_2 \cos\varphi_1 \sin\varphi_2 \vec{e}_5 + \cos\theta_1 \sin\theta_2 \sin\varphi_1 \vec{e}_6$$
$$+ \sin\theta_1 \cos\varphi_1 \cos\varphi_2 \vec{e}_7 + \sin\theta_1 \cos\varphi_1 \sin\varphi_2 \vec{e}_8 + \sin\theta_1 \sin\varphi_1 \vec{e}_9,$$
$$\partial\mathfrak{C}_2 = \cos\theta_1 \cos\theta_2 \cos\varphi_1 \cos\varphi_2 \vec{e}_{10} + \cos\theta_1 \cos\theta_2 \cos\varphi_1 \sin\varphi_2 \vec{e}_{11} + \cos\theta_1 \cos\theta_2 \sin\varphi_1 \vec{e}_{12}$$
$$+ \cos\theta_1 \sin\theta_2 \cos\varphi_1 \cos\varphi_2 \vec{e}_{13} + \cos\theta_1 \sin\theta_2 \cos\varphi_1 \sin\varphi_2 \vec{e}_{14} + \cos\theta_1 \sin\theta_2 \sin\varphi_1 \vec{e}_{15}$$
$$+ \sin\theta_1 \cos\varphi_1 \cos\varphi_2 \vec{e}_{16} + \sin\theta_1 \cos\varphi_1 \sin\varphi_2 \vec{e}_{17} + \sin\theta_1 \sin\varphi_1 \vec{e}_{18},$$
$$\partial\mathfrak{C}_3 = \cos\theta_1 \cos\theta_2 \cos\varphi_1 \cos\varphi_2 \vec{e}_{19} + \cos\theta_1 \cos\theta_2 \cos\varphi_1 \sin\varphi_2 \vec{e}_{20} + \cos\theta_1 \cos\theta_2 \sin\varphi_1 \vec{e}_{21}$$
$$+ \cos\theta_1 \sin\theta_2 \cos\varphi_1 \cos\varphi_2 \vec{e}_{22} + \cos\theta_1 \sin\theta_2 \cos\varphi_1 \sin\varphi_2 \vec{e}_{23} + \cos\theta_1 \sin\theta_2 \sin\varphi_1 \vec{e}_{24}$$
$$+ \sin\theta_1 \cos\varphi_1 \cos\varphi_2 \vec{e}_{25} + \sin\theta_1 \cos\varphi_1 \sin\varphi_2 \vec{e}_{26} + \sin\theta_1 \sin\varphi_1 \vec{e}_{27}. \tag{B5}$$

Obviously, any point in $\mathfrak{C}$ can be expressed as a convex combination of three points, each belonging to the regions $\mathfrak{C}_1, \mathfrak{C}_2$, and $\mathfrak{C}_3$, respectively. Furthermore, any point in $\mathfrak{C}_1$ ($\mathfrak{C}_2$ or $\mathfrak{C}_3$) is a convex combination of two points located on the boundary of $\mathfrak{C}_1$ ($\mathfrak{C}_2$ or $\mathfrak{C}_3$). Denote $\vec{0}$ the zero nine-dimensional vector. Define

$$\vec{h}_j = (\cos\theta_{1j} \cos\theta_{2j} \cos\varphi_{1j} \cos\varphi_{2j}, \cos\theta_{1j} \cos\theta_{2j} \cos\varphi_{1j} \sin\varphi_{2j}, \cos\theta_{1j} \cos\theta_{2j} \sin\varphi_{1j},$$
$$\cos\theta_{1j} \sin\theta_{2j} \cos\varphi_{1j} \cos\varphi_{2j}, \cos\theta_{1j} \sin\theta_{2j} \cos\varphi_{1j} \sin\varphi_{2j}, \cos\theta_{1j} \sin\theta_{2j} \sin\varphi_{1j},$$
$$\sin\theta_{1j} \cos\varphi_{1j} \cos\varphi_{2j}, \sin\theta_{1j} \cos\varphi_{1j} \sin\varphi_{2j}, \sin\theta_{1j} \sin\varphi_{1j}), \tag{B6}$$

where $k = 1, 2, 3, 4, 5$, and 6 denotes different values of the parameters $\theta_1, \theta_2, \varphi_1$, and $\varphi_2$. Then, any point in $\mathfrak{C}$ can be written as

$$\vec{v} = q_1(\vec{h_1}, \vec{0}, \vec{0}) + q_2(\vec{h_2}, \vec{0}, \vec{0}) + q_3(\vec{0}, \vec{h_3}, \vec{0})$$
$$+ q_4(\vec{0}, \vec{h_4}, \vec{0}) + q_5(\vec{0}, \vec{0}, \vec{h_5}) + q_6(\vec{0}, \vec{0}, \vec{h_6}), \tag{B7}$$

where $\vec{0}$ is the nine-dimensional zero vector. One easily proves that $(q_1 h_1 + q_2 h_2)^2 \leqslant (q_1 + q_2)^2$, $(q_3 h_3 + q_4 h_4)^2 \leqslant (q_3 + q_4)^2$, and $(q_5 h_5 + q_6 h_6)^2 \leqslant (q_5 + q_6)^2$. Therefore, we have

$$\sqrt{\sum_{i=1}^{9} v_i^2} + \sqrt{\sum_{i=10}^{18} v_i^2} + \sqrt{\sum_{i=19}^{27} v_i^2} \leqslant \sum_{j=1}^{6} q_j = 1. \tag{B8}$$

Substituting the inequality back to the one in the original basis, we derive the NLHS steering inequality as follows,

$$\sqrt{\sum_{x,y=1}^{3} \langle A_x(C^2 + C^3) B_y \rangle^2} + \sqrt{\sum_{x,y=1}^{3} \langle A_x(C^1 - C^2) B_y \rangle^2} + \sqrt{\sum_{x,y=1}^{3} \langle A_x(C^1 - C^3) B_y \rangle^2} \leqslant 2. \tag{B9}$$

∎

## APPENDIX C: EXAMPLES OF NETWORK STEERING DETECTABLE BY NCHSH-LIKE INEQUALITY BUT NOT BY ENTANGLEMENT CRITERIA

Here, we present three examples of network steering, which can be detected by our NCHSH-like inequality but cannot be detected by entanglement criteria.

*Example C1.*

$$\rho^{AC} = \begin{pmatrix} 0.1176 + 0.0000i & 0.1338 + 0.1784i & 0.0352 - 0.0790i & -0.0147 - 0.0696i \\ 0.1338 - 0.1784i & 0.6046 + 0.0000i & -0.1272 - 0.1190i & -0.2105 - 0.0627i \\ 0.0352 + 0.0790i & -0.1272 + 0.1190i & 0.1589 - 0.0000i & 0.0204 + 0.0083i \\ -0.0147 + 0.0696i & -0.2105 + 0.0627i & 0.0204 - 0.0083i & 0.1189 - 0.0000i \end{pmatrix}, \tag{C1}$$

$$\rho^{BC'} = \begin{pmatrix} 0.1124 + 0.0000i & -0.1549 + 0.0708i & -0.0984 - 0.0210i & 0.1142 + 0.0472i \\ -0.1549 - 0.0708i & 0.5344 + 0.0000i & 0.1369 + 0.0400i & -0.2966 - 0.1985i \\ -0.0984 + 0.0210i & 0.1369 - 0.0400i & 0.1002 - 0.0000i & -0.1064 - 0.0544i \\ 0.1142 - 0.0472i & -0.2966 + 0.1985i & -0.1064 + 0.0544i & 0.2530 + 0.0000i \end{pmatrix}. \tag{C2}$$

*PPT criterion.* After Charlie performs a generalized EJM with $\theta = \pi/2$, four conditional states and their normalized eigenvalues are given by

$$\rho_1 = \begin{pmatrix} 0.1405 + 0.0000i & -0.1411 - 0.0764i & -0.0723 - 0.0556i & 0.0546 + 0.0921i \\ -0.1411 + 0.0764i & 0.3227 + 0.0000i & 0.1063 + 0.0066i & -0.1932 - 0.0774i \\ -0.0723 + 0.0556i & 0.1063 - 0.0066i & 0.1510 + 0.0000i & -0.1674 - 0.0203i \\ 0.0546 - 0.0921i & -0.1932 + 0.0774i & -0.1674 + 0.0203i & 0.3858 - 0.0000i \end{pmatrix},$$

$$\lambda_{11} = 0.7131, \quad \lambda_{12} = 0.1992, \quad \lambda_{13} = 0.0257, \quad \lambda_{14} = 0.0619,$$

$$\rho_2 = \begin{pmatrix} 0.1585 + 0.0000i & -0.2143 + 0.0273i & -0.0388 - 0.0330i & 0.0643 + 0.0387i \\ -0.2143 - 0.0273i & 0.6694 - 0.0000i & 0.0431 + 0.0517i & -0.1767 - 0.1446i \\ -0.0388 + 0.0330i & 0.0431 - 0.0517i & 0.0329 + 0.0000i & -0.0442 + 0.0076i \\ 0.0643 - 0.0387i & -0.1767 + 0.1446i & -0.0442 - 0.0076i & 0.1391 + 0.0000i \end{pmatrix},$$

$$\lambda_{21} = 0.8420, \quad \lambda_{22} = 0.0071, \quad \lambda_{23} = 0.0907, \quad \lambda_{24} = 0.0601,$$

$$\rho_3 = \begin{pmatrix} 0.1899 + 0.0000i & -0.1735 + 0.0842i & -0.0165 - 0.0036i & -0.0400 - 0.0310i \\ -0.1735 - 0.0842i & 0.4270 - 0.0000i & 0.0301 + 0.0232i & 0.0062 + 0.0550i \\ -0.0165 + 0.0036i & 0.0301 - 0.0232i & 0.0725 + 0.0000i & -0.0866 + 0.0507i \\ -0.0400 + 0.0310i & 0.0062 - 0.0550i & -0.0866 - 0.0507i & 0.3106 + 0.0000i \end{pmatrix},$$

$$\lambda_{31} = 0.5517, \quad \lambda_{32} = 0.3441, \quad \lambda_{33} = 0.0765, \quad \lambda_{34} = 0.0277,$$

$$\rho_4 = \begin{pmatrix} 0.0819 + 0.0000i & -0.0451 - 0.0657i & 0.0181 + 0.0134i & -0.0119 - 0.0381i \\ -0.0451 + 0.0657i & 0.2201 + 0.0000i & -0.0566 + 0.0074i & 0.1676 - 0.0628i \\ 0.0181 - 0.0134i & -0.0566 - 0.0074i & 0.1255 - 0.0000i & -0.1549 - 0.0378i \\ -0.0119 + 0.0381i & 0.1676 + 0.0628i & -0.1549 + 0.0378i & 0.5726 - 0.0000i \end{pmatrix},$$

$$\lambda_{21} = 0.6969, \quad \lambda_{22} = 0.2156, \quad \lambda_{23} = 0.0220, \quad \lambda_{24} = 0.0654.$$

Hence, the PPT criterion fails to detect network steering.

*NCHSH-like inequality.* Alice and Bob perform $A_1 = B_1 = \sigma_x$, $A_2 = B_2 = \sigma_y$, and $A_3 = B_3 = \sigma_z$, and Charlie performs a generalized EJM with $\theta = \pi/2$. Then we obtain

$$\sqrt{\sum_{x,y=1}^{3} \text{Tr}\{[A_x(C^2 + C^3)B_y](\rho^{AC} \otimes \rho^{BC'})\}^2} + \sqrt{\sum_{x,y=1}^{3} \text{Tr}\{[A_x(C^1 - C^2)B_y](\rho^{AC} \otimes \rho^{BC'})\}^2}$$
$$+ \sqrt{\sum_{x,y=1}^{3} \text{Tr}\{[A_x(C^1 - C^3)B_y](\rho^{AC} \otimes \rho^{BC'})\}^2} = 2.9628. \quad (C3)$$

Hence, our NCHSH-like inequality succeeds to detect network steering.

*Example C2.*

$$\rho^{AC} = \begin{pmatrix} 0.1269 + 0.0000i & 0.0213 + 0.0356i & -0.0266 + 0.0479i & -0.0796 - 0.1103i \\ 0.0213 - 0.0356i & 0.2567 + 0.0000i & 0.0960 + 0.2936i & -0.0532 + 0.1576i \\ -0.0266 - 0.0479i & 0.0960 - 0.2936i & 0.3728 + 0.0000i & 0.1478 + 0.1185i \\ -0.0796 + 0.1103i & -0.0532 - 0.1576i & 0.1478 - 0.1185i & 0.2436 + 0.0000i \end{pmatrix}, \quad (C4)$$

$$\rho^{BC'} = \begin{pmatrix} 0.3889 + 0.0000i & -0.1272 + 0.0100i & -0.0625 - 0.0103i & 0.0632 + 0.1752i \\ -0.1272 - 0.0100i & 0.1718 - 0.0000i & 0.2026 + 0.0916i & -0.0531 - 0.1037i \\ -0.0625 + 0.0103i & 0.2026 - 0.0916i & 0.3242 - 0.0000i & -0.0965 - 0.0647i \\ 0.0632 - 0.1752i & -0.0531 + 0.1037i & -0.0965 + 0.0647i & 0.1151 + 0.0000i \end{pmatrix}. \quad (C5)$$

*Example C3.*

$$\rho^{AC} = \begin{pmatrix} 0.1176 + 0.0000i & 0.1338 + 0.1784i & 0.0352 - 0.0790i & -0.0147 - 0.0696i \\ 0.2496 + 0.0000i & 0.3208 - 0.0056i & -0.0938 - 0.0680i & -0.1746 - 0.0558i \\ 0.3208 + 0.0056i & 0.4646 + 0.0000i & -0.0937 - 0.0276i & -0.2404 - 0.0930i \\ -0.0938 + 0.0680i & -0.0937 + 0.0276i & 0.1397 - 0.0000i & 0.0518 - 0.0146i \\ -0.1746 + 0.0558i & -0.2404 + 0.0930i & 0.0518 + 0.0146i & 0.1461 + 0.0000i \end{pmatrix}, \quad (C6)$$

$$\rho^{BC'} = \begin{pmatrix} 0.3159 - 0.0000i & -0.1171 - 0.2229i & 0.2041 - 0.1446i & -0.1499 + 0.0392i \\ -0.1171 + 0.2229i & 0.2134 - 0.0000i & 0.0650 + 0.2020i & 0.0391 - 0.0909i \\ 0.2041 + 0.1446i & 0.0650 - 0.2020i & 0.3170 + 0.0000i & -0.0707 + 0.0421i \\ -0.1499 - 0.0392i & 0.0391 + 0.0909i & -0.0707 - 0.0421i & 0.1537 - 0.0000i \end{pmatrix}. \quad (C7)$$

### APPENDIX D: PROOF OF THEOREM 2

In the case of the dual-node repeater network admitting an NLHS model, the expectations are given by

$$\langle A_x C^k D^l B_y \rangle = \sum_{a,c,d,b} ac^k d^l b p(a, c, d, b|x, y) = \sum_{c,d,\xi,\zeta} p(c, d, \xi, \zeta) c^k d^l \langle A_x \rangle_{|\phi_\xi\rangle\langle\phi_\xi|} \langle B_y \rangle_{|\psi_\zeta\rangle\langle\psi_\zeta|}. \quad (D1)$$

Then, all of the 81 expectations are presented in Table II.

Obviously, the 1st, 2nd, 3rd, 5th, 6th, 7th, 9th, 10th, and 11th columns form the maximal linearly independent set among the above 16 column vectors. Hence, if a dual-node repeater network admits an NLHS model under the specified measurements, the corresponding 81 expectations must be a convex combination of such nine columns for some values of $\theta_1, \theta_2, \varphi_1$, and $\varphi_2$. Thus, if we define $\{\mathfrak{C}_i\}_{i=1}^{16}$ as the convex hull of each column, then the 81 expectations of NLHS model model reside within the convex hull:

$$\mathfrak{C} = \text{convex}\left(\bigcup_{i=1}^{16} \mathfrak{C}_i\right).$$

TABLE II. Expectations of the NLHS model in dual-node repeater networks.

| Expectations | $c = (1, 1, 1)$<br>$d = (1, 1, 1)$ | $c = (1, 1, 1)$<br>$d = (1, -1, -1)$ | $c = (1, 1, 1)$<br>$d = (-1, 1, -1)$ | $c = (1, 1, 1)$<br>$d = (-1, -1, 1)$ | $c = (1, -1, -1)$<br>$d = (1, 1, 1)$ | $c = (1, -1, -1)$<br>$d = (1, -1, -1)$ |
|---|---|---|---|---|---|---|
| $\langle A_x C^1 D^1 B_y \rangle$ | $\langle A_x \rangle \langle B_y \rangle$ | $\langle A_x \rangle \langle B_y \rangle$ | $-\langle A_x \rangle \langle B_y \rangle$ | $-\langle A_x \rangle \langle B_y \rangle$ | $\langle A_x \rangle \langle B_y \rangle$ | $\langle A_x \rangle \langle B_y \rangle$ |
| $\langle A_x C^1 D^2 B_y \rangle$ | $\langle A_x \rangle \langle B_y \rangle$ | $-\langle A_x \rangle \langle B_y \rangle$ | $\langle A_x \rangle \langle B_y \rangle$ | $-\langle A_x \rangle \langle B_y \rangle$ | $\langle A_x \rangle \langle B_y \rangle$ | $-\langle A_x \rangle \langle B_y \rangle$ |
| $\langle A_x C^1 D^3 B_y \rangle$ | $\langle A_x \rangle \langle B_y \rangle$ | $-\langle A_x \rangle \langle B_y \rangle$ | $-\langle A_x \rangle \langle B_y \rangle$ | $\langle A_x \rangle \langle B_y \rangle$ | $\langle A_x \rangle \langle B_y \rangle$ | $-\langle A_x \rangle \langle B_y \rangle$ |
| $\langle A_x C^2 D^1 B_y \rangle$ | $\langle A_x \rangle \langle B_y \rangle$ | $\langle A_x \rangle \langle B_y \rangle$ | $-\langle A_x \rangle \langle B_y \rangle$ | $-\langle A_x \rangle \langle B_y \rangle$ | $-\langle A_x \rangle \langle B_y \rangle$ | $-\langle A_x \rangle \langle B_y \rangle$ |
| $\langle A_x C^2 D^2 B_y \rangle$ | $\langle A_x \rangle \langle B_y \rangle$ | $-\langle A_x \rangle \langle B_y \rangle$ | $\langle A_x \rangle \langle B_y \rangle$ | $-\langle A_x \rangle \langle B_y \rangle$ | $-\langle A_x \rangle \langle B_y \rangle$ | $\langle A_x \rangle \langle B_y \rangle$ |
| $\langle A_x C^2 D^3 B_y \rangle$ | $\langle A_x \rangle \langle B_y \rangle$ | $-\langle A_x \rangle \langle B_y \rangle$ | $-\langle A_x \rangle \langle B_y \rangle$ | $\langle A_x \rangle \langle B_y \rangle$ | $-\langle A_x \rangle \langle B_y \rangle$ | $\langle A_x \rangle \langle B_y \rangle$ |
| $\langle A_x C^3 D^1 B_y \rangle$ | $\langle A_x \rangle \langle B_y \rangle$ | $\langle A_x \rangle \langle B_y \rangle$ | $-\langle A_x \rangle \langle B_y \rangle$ | $-\langle A_x \rangle \langle B_y \rangle$ | $-\langle A_x \rangle \langle B_y \rangle$ | $-\langle A_x \rangle \langle B_y \rangle$ |
| $\langle A_x C^3 D^2 B_y \rangle$ | $\langle A_x \rangle \langle B_y \rangle$ | $-\langle A_x \rangle \langle B_y \rangle$ | $\langle A_x \rangle \langle B_y \rangle$ | $-\langle A_x \rangle \langle B_y \rangle$ | $-\langle A_x \rangle \langle B_y \rangle$ | $\langle A_x \rangle \langle B_y \rangle$ |
| $\langle A_x C^3 D^3 B_y \rangle$ | $\langle A_x \rangle \langle B_y \rangle$ | $-\langle A_x \rangle \langle B_y \rangle$ | $-\langle A_x \rangle \langle B_y \rangle$ | $\langle A_x \rangle \langle B_y \rangle$ | $-\langle A_x \rangle \langle B_y \rangle$ | $\langle A_x \rangle \langle B_y \rangle$ |
| Expectations | $c = (1, -1, -1)$<br>$d = (-1, 1, -1)$ | $c = (1, -1, -1)$<br>$d = (-1, -1, 1)$ | $c = (-1, 1, -1)$<br>$d = (1, 1, 1)$ | $c = (-1, 1, -1)$<br>$d = (1, -1, -1)$ | $c = (-1, 1, -1)$<br>$d = (-1, 1, -1)$ | $c = (-1, 1, -1)$<br>$d = (-1, -1, 1)$ |
| $\langle A_x C^1 D^1 B_y \rangle$ | $-\langle A_x \rangle \langle B_y \rangle$ | $-\langle A_x \rangle \langle B_y \rangle$ | $-\langle A_x \rangle \langle B_y \rangle$ | $-\langle A_x \rangle \langle B_y \rangle$ | $\langle A_x \rangle \langle B_y \rangle$ | $\langle A_x \rangle \langle B_y \rangle$ |
| $\langle A_x C^1 D^2 B_y \rangle$ | $\langle A_x \rangle \langle B_y \rangle$ | $-\langle A_x \rangle \langle B_y \rangle$ | $-\langle A_x \rangle \langle B_y \rangle$ | $\langle A_x \rangle \langle B_y \rangle$ | $-\langle A_x \rangle \langle B_y \rangle$ | $\langle A_x \rangle \langle B_y \rangle$ |
| $\langle A_x C^1 D^3 B_y \rangle$ | $-\langle A_x \rangle \langle B_y \rangle$ | $\langle A_x \rangle \langle B_y \rangle$ | $-\langle A_x \rangle \langle B_y \rangle$ | $\langle A_x \rangle \langle B_y \rangle$ | $\langle A_x \rangle \langle B_y \rangle$ | $-\langle A_x \rangle \langle B_y \rangle$ |
| $\langle A_x C^2 D^1 B_y \rangle$ | $\langle A_x \rangle \langle B_y \rangle$ | $\langle A_x \rangle \langle B_y \rangle$ | $\langle A_x \rangle \langle B_y \rangle$ | $\langle A_x \rangle \langle B_y \rangle$ | $-\langle A_x \rangle \langle B_y \rangle$ | $-\langle A_x \rangle \langle B_y \rangle$ |
| $\langle A_x C^2 D^2 B_y \rangle$ | $-\langle A_x \rangle \langle B_y \rangle$ | $\langle A_x \rangle \langle B_y \rangle$ | $\langle A_x \rangle \langle B_y \rangle$ | $-\langle A_x \rangle \langle B_y \rangle$ | $\langle A_x \rangle \langle B_y \rangle$ | $-\langle A_x \rangle \langle B_y \rangle$ |
| $\langle A_x C^2 D^3 B_y \rangle$ | $\langle A_x \rangle \langle B_y \rangle$ | $-\langle A_x \rangle \langle B_y \rangle$ | $\langle A_x \rangle \langle B_y \rangle$ | $-\langle A_x \rangle \langle B_y \rangle$ | $-\langle A_x \rangle \langle B_y \rangle$ | $\langle A_x \rangle \langle B_y \rangle$ |
| $\langle A_x C^3 D^1 B_y \rangle$ | $\langle A_x \rangle \langle B_y \rangle$ | $\langle A_x \rangle \langle B_y \rangle$ | $-\langle A_x \rangle \langle B_y \rangle$ | $-\langle A_x \rangle \langle B_y \rangle$ | $\langle A_x \rangle \langle B_y \rangle$ | $\langle A_x \rangle \langle B_y \rangle$ |
| $\langle A_x C^3 D^2 B_y \rangle$ | $-\langle A_x \rangle \langle B_y \rangle$ | $\langle A_x \rangle \langle B_y \rangle$ | $-\langle A_x \rangle \langle B_y \rangle$ | $\langle A_x \rangle \langle B_y \rangle$ | $-\langle A_x \rangle \langle B_y \rangle$ | $\langle A_x \rangle \langle B_y \rangle$ |
| $\langle A_x C^3 D^3 B_y \rangle$ | $\langle A_x \rangle \langle B_y \rangle$ | $-\langle A_x \rangle \langle B_y \rangle$ | $-\langle A_x \rangle \langle B_y \rangle$ | $\langle A_x \rangle \langle B_y \rangle$ | $\langle A_x \rangle \langle B_y \rangle$ | $-\langle A_x \rangle \langle B_y \rangle$ |
| Expectations | $c = (-1, -1, 1)$<br>$d = (1, 1, 1)$ | $c = (-1, -1, 1)$<br>$d = (1, -1, -1)$ | $c = (-1, -1, 1)$<br>$d = (-1, 1, -1)$ | $c = (-1, -1, 1)$<br>$d = (-1, -1, 1)$ | | |
| $\langle A_x C^1 D^1 B_y \rangle$ | $-\langle A_x \rangle \langle B_y \rangle$ | $-\langle A_x \rangle \langle B_y \rangle$ | $\langle A_x \rangle \langle B_y \rangle$ | $\langle A_x \rangle \langle B_y \rangle$ | | |
| $\langle A_x C^1 D^2 B_y \rangle$ | $-\langle A_x \rangle \langle B_y \rangle$ | $\langle A_x \rangle \langle B_y \rangle$ | $-\langle A_x \rangle \langle B_y \rangle$ | $\langle A_x \rangle \langle B_y \rangle$ | | |
| $\langle A_x C^1 D^3 B_y \rangle$ | $-\langle A_x \rangle \langle B_y \rangle$ | $\langle A_x \rangle \langle B_y \rangle$ | $\langle A_x \rangle \langle B_y \rangle$ | $-\langle A_x \rangle \langle B_y \rangle$ | | |
| $\langle A_x C^2 D^1 B_y \rangle$ | $-\langle A_x \rangle \langle B_y \rangle$ | $-\langle A_x \rangle \langle B_y \rangle$ | $\langle A_x \rangle \langle B_y \rangle$ | $\langle A_x \rangle \langle B_y \rangle$ | | |
| $\langle A_x C^2 D^2 B_y \rangle$ | $-\langle A_x \rangle \langle B_y \rangle$ | $\langle A_x \rangle \langle B_y \rangle$ | $-\langle A_x \rangle \langle B_y \rangle$ | $\langle A_x \rangle \langle B_y \rangle$ | | |
| $\langle A_x C^2 D^3 B_y \rangle$ | $-\langle A_x \rangle \langle B_y \rangle$ | $\langle A_x \rangle \langle B_y \rangle$ | $\langle A_x \rangle \langle B_y \rangle$ | $-\langle A_x \rangle \langle B_y \rangle$ | | |
| $\langle A_x C^3 D^1 B_y \rangle$ | $\langle A_x \rangle \langle B_y \rangle$ | $\langle A_x \rangle \langle B_y \rangle$ | $-\langle A_x \rangle \langle B_y \rangle$ | $-\langle A_x \rangle \langle B_y \rangle$ | | |
| $\langle A_x C^3 D^2 B_y \rangle$ | $\langle A_x \rangle \langle B_y \rangle$ | $-\langle A_x \rangle \langle B_y \rangle$ | $\langle A_x \rangle \langle B_y \rangle$ | $-\langle A_x \rangle \langle B_y \rangle$ | | |
| $\langle A_x C^3 D^3 B_y \rangle$ | $\langle A_x \rangle \langle B_y \rangle$ | $-\langle A_x \rangle \langle B_y \rangle$ | $-\langle A_x \rangle \langle B_y \rangle$ | $\langle A_x \rangle \langle B_y \rangle$ | | |

In order to better describe the region $\mathfrak{C}$, we define a new orthogonal basis in $\mathbb{R}^{81}$. If we set $f_i$ as a nine-dimensional vector where the *i*th component is 1 and all other components are 0, then the new basis is defined as

$$
\begin{aligned}
e_i &= (f_i, f_i, f_i, f_i, f_i, f_i, f_i, f_i, f_i), & e_{i+9} &= (f_i, -f_i, -f_i, f_i, -f_i, -f_i, f_i, -f_i, -f_i), \\
e_{i+18} &= (-f_i, f_i, -f_i, -f_i, f_i, -f_i, -f_i, f_i, -f_i), & e_{i+27} &= (f_i, f_i, f_i, -f_i, -f_i, -f_i, -f_i, -f_i, -f_i), \\
e_{i+36} &= (f_i, -f_i, -f_i, -f_i, f_i, f_i, -f_i, f_i, f_i), & e_{i+45} &= (-f_i, f_i, -f_i, f_i, -f_i, f_i, f_i, -f_i, f_i), \\
e_{i+54} &= (-f_i, -f_i, f_i, f_i, f_i, -f_i, -f_i, -f_i, f_i), & e_{i+63} &= (-f_i, f_i, f_i, f_i, -f_i, f_i, -f_i, -f_i, f_i), \\
e_{i+72} &= (f_i, -f_i, f_i, -f_i, f_i, -f_i, -f_i, -f_i, f_i),
\end{aligned}
\tag{D2}
$$

where $1 \leqslant i \leqslant 9$. We set $\vec{h} = (\cos\theta_1 \cos\theta_2, \cos\theta_1 \sin\theta_2, \sin\theta_1) \otimes (\cos\varphi_1 \cos\varphi_2, \cos\varphi_1 \sin\varphi_2, \sin\varphi_1)$ and $\{\vec{h_j}\}_{j=1}^{18}$ as vectors for some fixed parameters $\theta_1, \theta_2, \varphi_1$, and $\varphi_2$, and then the boundaries of $\mathfrak{C}_1, \mathfrak{C}_2, \mathfrak{C}_3, \mathfrak{C}_5, \mathfrak{C}_6, \mathfrak{C}_7, \mathfrak{C}_9, \mathfrak{C}_{10}$, and $\mathfrak{C}_{11}$ are of the

following forms:

$$\partial\mathfrak{C}_1 = h \cdot (e_1, e_2, e_3, e_4, e_5, e_6, e_7, e_8, e_9), \quad \partial\mathfrak{C}_2 = h \cdot (e_{10}, e_{11}, e_{12}, e_{13}, e_{14}, e_{15}, e_{16}, e_{17}, e_{18}),$$
$$\partial\mathfrak{C}_3 = h \cdot (e_{19}, e_{20}, e_{21}, e_{22}, e_{23}, e_{24}, e_{25}, e_{26}, e_{27}), \quad \partial\mathfrak{C}_4 = h \cdot (e_{28}, e_{29}, e_{30}, e_{31}, e_{32}, e_{33}, e_{34}, e_{35}, e_{36}),$$
$$\partial\mathfrak{C}_5 = h \cdot (e_{37}, e_{38}, e_{39}, e_{40}, e_{41}, e_{42}, e_{43}, e_{44}, e_{45}), \quad \partial\mathfrak{C}_6 = h \cdot (e_{46}, e_{47}, e_{48}, e_{49}, e_{50}, e_{51}, e_{52}, e_{53}, e_{54}),$$
$$\partial\mathfrak{C}_7 = h \cdot (e_{55}, e_{56}, e_{57}, e_{58}, e_{59}, e_{60}, e_{61}, e_{62}, e_{63}), \quad \partial\mathfrak{C}_8 = h \cdot (e_{64}, e_{65}, e_{66}, e_{67}, e_{68}, e_{69}, e_{70}, e_{71}, e_{72}),$$
$$\partial\mathfrak{C}_9 = h \cdot (e_{73}, e_{74}, e_{75}, e_{76}, e_{77}, e_{78}, e_{79}, e_{80}, e_{81}).$$

Any point in $\mathfrak{C}$ can be written as

$$\vec{v} = \sum_{k=1}^{2} q_j(\vec{h_j}, \vec{0}, \vec{0}, \vec{0}, \vec{0}, \vec{0}, \vec{0}, \vec{0}, \vec{0}) + \sum_{k=3}^{4} q_j(\vec{0}, \vec{h_j}, \vec{0}, \vec{0}, \vec{0}, \vec{0}, \vec{0}, \vec{0}, \vec{0})$$
$$+ \sum_{k=5}^{6} q_j(\vec{0}, \vec{0}, \vec{h_j}, \vec{0}, \vec{0}, \vec{0}, \vec{0}, \vec{0}, \vec{0}) + \sum_{k=7}^{8} q_j(\vec{0}, \vec{0}, \vec{0}, \vec{h_j}, \vec{0}, \vec{0}, \vec{0}, \vec{0}, \vec{0})$$
$$+ \sum_{k=9}^{10} q_j(\vec{0}, \vec{0}, \vec{0}, \vec{0}, \vec{h_j}, \vec{0}, \vec{0}, \vec{0}, \vec{0}) + \sum_{k=11}^{12} q_j(\vec{0}, \vec{0}, \vec{0}, \vec{0}, \vec{0}, \vec{h_j}, \vec{0}, \vec{0}, \vec{0})$$
$$+ \sum_{k=13}^{14} q_j(\vec{0}, \vec{0}, \vec{0}, \vec{0}, \vec{0}, \vec{0}, \vec{h_j}, \vec{0}, \vec{0}) + \sum_{k=15}^{16} q_j(\vec{0}, \vec{0}, \vec{0}, \vec{0}, \vec{0}, \vec{0}, \vec{0}, \vec{h_j}, \vec{0})$$
$$+ \sum_{k=17}^{18} q_j(\vec{0}, \vec{0}, \vec{0}, \vec{0}, \vec{0}, \vec{0}, \vec{0}, \vec{0}, \vec{h_j}). \tag{D3}$$

Similarly to Eq. (B8), we obtain

$$\sum_{r=0}^{8} \sqrt{\sum_{i=1}^{9} v_{i+9r}^2} \leqslant \sum_{k=1}^{18} q_j = 1. \tag{D4}$$

Substituting the inequality back to the one in the original basis, we derive the following dual-node NCHSH-like inequality:

$$\sum_{\substack{1 \leqslant m < n \leqslant 3 \\ 1 \leqslant p < q \leqslant 3}} \sqrt{\sum_{x,y=1}^{3} \langle A_x [C^m + (-1)^m C^n][D^p + (-1)^p D^q] B_y \rangle^2} \leqslant 4. \tag{D5}$$